\documentstyle[12pt]{article}
\def\gsim{\stackrel{>}{\sim}}
\def\lsim{\stackrel{<}{\sim}}
\def\beq{\begin{equation}}
\def\eeq{\end{equation}}

\def\half{\frac{1}{2}}
\def\ie{i.\ e.\ }
\def\etal{et al.\ }
\begin{document}
\begin{flushright}
VAND--TH-94--14\\
July 1994
\end{flushright}
\begin{center}
\Large
{\bf Higgs Mass Bounds Separate Models of Electroweak Symmetry Breaking}
\end{center}
\normalsize
\bigskip
\begin{center}
{\large Marco A.  D\'\i az, Tonnis A. ter Veldhuis} and
{\large Thomas J. Weiler}\\
{\sl Department of Physics \& Astronomy\\
Vanderbilt University\\
Nashville, TN 37235, USA}\\
\vspace{2cm}
\end{center}
\centerline{ABSTRACT}
Vacuum stability implies a lower limit on the
mass of the higgs boson in the Standard Model (SM).
In contrast, an upper limit
on the lightest higgs mass can be calculated in supersymmetric (susy) models.
The main uncertainty in each limit is the value of the top mass,
which may now be fixed by the recent CDF result.
We study the possibility
that these bounds do not overlap, and find that\\
(i) a mass gap emerges at $m_t\sim 160$ GeV
between the SM and the Minimal Susy Standard Model (MSSM);
and between the SM and the Minimal plus Singlet Susy Model [(M+1)SSM] if
the independent scalar self--coupling of the latter is perturbatively small
or if the $\tan\beta$ parameter is large;
this gap widens with increasing $m_t$; \\
(ii) there is no overlap between the SM and
the MSSM bounds at even smaller values of $m_t$ for
the $\tan\beta$ value ($\sim 1$--2) preferred in
Supersymmetric Grand Unified Theories.\\
Thus, if the new top mass measurement remains valid, a
measurement of the first higgs mass will serve to exclude either the SM
or MSSM/(M+1)SSM higgs sectors.
In addition, we discuss the upper bound on the lightest higgs mass
in susy models with an extended higgs sector, and in models
with a strongly interacting higgs sector.
Finally, we comment on the discovery potential for
the lightest higgses in these models.
\vfill\eject
\section{Introduction}

The simplest and best motivated
possibilities for the electroweak symmetry breaking
sector are the single higgs doublet of the minimal Standard Model SM,
and the two higgs doublet sector of the Minimal Supersymmetric
Standard Model (MSSM).  Experimentally, very little is known about the
higgs sector of the electroweak model.  However, theoretically, quite a
lot of higgs physics has been calculated.
The electroweak symmetry--breaking scale is known:
the vacuum expectation value (vev) of the complex higgs field $\Phi$
is $<0|\Phi|0>=v_{SM}/\sqrt{2}=175$ GeV.
This value is remarkably close to
the probable top quark mass of $174\pm10^{+13}_{-12}$ GeV
(very consistent with the SM prediction of
$m_t=164\pm25$ GeV inferred from precision electroweak data \cite{ewgroup})
announced recently by the CDF collaboration at Fermilab \cite{CDF}.
Higgs mass bounds have been calculated, including loop corrections.
One aspect of the mass bounds \cite{kp} which we quantify
in this paper is the following:
inputing the CDF value for the top mass into quantum loop
corrections for the symmetry-- breaking higgs sector leads to mutually
exclusive, reliable bounds on the SM higgs mass and
on the lightest MSSM higgs mass.  From
this we infer that
{\it if the CDF value for $m_t$ is
verified, then the first higgs mass measurement
will rule out one of the two main contenders (SM vs.\ MSSM)
for \underline{the} electroweak theory, independent of any other measurement.}

There is another point to be made here.
It is known that the Feynman rules connecting the lightest higgs in the MSSM
to ordinary matter become, in the limit
where the ``other'' higgs masses (these are $m_A, m_H$, and $m_{H^{\pm}}$,
defined in section 3) are taken to infinity,
exactly the SM Feynman rules \cite{HHG}.  When the masses are taken
large compared to $M_Z$, of the order of
a TeV, for example, the lightest MSSM higgs behaves very much
like the SM higgs in its production channels and decay modes \cite{cpr1};
the only difference, a vestige of the underlying supersymmetry,
is that the constrained higgs self coupling requires the MSSM higgs to be
light,
whereas SM vacuum stability requires the SM higgs to be heavy.  Thus,
{\it there may be no discernible difference between the lightest
MSSM higgs and the SM higgs, except for their allowed mass values.}
We demonstrate these allowed mass values in our Figures 1 and 2.
Furthermore, the mass of the lightest MSSM higgs rises toward its upper bound
as the ``other'' higgs masses are increased
\footnote
{The saturation of the MSSM upper bound with increasing ``other'' higgs masses
is well known in tree--level relations (the bound $m_h \leq M_Z|\cos(2\beta)|$
approaches an equality as higgs masses increase) \cite{HHG}.
The MSSM upper bound still saturates
with increasing ``other'' higgs masses even when one--loop corrections
are included.
}.
Thus, for masses in the region where the SM lower bound
and the MSSM upper bound overlap,
the SM higgs and the lightest MSSM higgs may not be distinguishable by
branching ratio or width measurements.
Only if the two bounds are separated by a gap is this ambiguity avoided.

In the SM and even in supersymmetric (susy) models the main
uncertainty in radiative corrections is the value of the top mass.
If the CDF announcement is confirmed, this main uncertainty is eliminated.
{\it The radiatively corrected observable
most sensitive to the value
of the top mass is the mass of the
lightest higgs particle in susy models {\rm \cite{mt4}}}:
for large top mass, the top and scalar--top ($\tilde{t}$) loops dominate
all other loop corrections, and
{\it the light higgs mass-squared grows as }
$m_t^4 \ln(m_{\tilde{t}}/m_t)$.
\footnote
{It is not hard to understand this fourth power dependence; the contribution
of the top loop to the SM higgs self energy also scales as $m_t^4$.  However,
in the SM the higgs mass is a free parameter at tree--level, and so any
radiative correction to the SM higgs mass is not measurable.  In contrast,
in the MSSM the lightest higgs mass at tree--level is fixed by other
observables, and so the finite renormalization is measurable.
}
We quantify this large correction in section 3.

In addition to contrasting the MSSM with the SM, we also consider
in section 4
supersymmetric models with a non-standard Higgs sector, in particular
the Minimal--plus--Singlet Susy
Standard Model [(M+1)SSM] containing an additional
SU(2) singlet, and a gauged non--linear sigma model.
A discussion of supersymmetric grand unified theories (susy GUTs)
is put forth in section 5; susy GUTs impose additional constraints on the
low energy MSSM, leading to a lower upper bound on the lightest higgs mass.
The discovery potential for the higgs boson in analyzed in section 6, and
conclusions are presented in section 7.

\section{Standard model vacuum stability bound}

Recently it has been shown that when the newly reported
value of the top mass is input
into the effective potential for the SM higgs field, the
broken--symmetry potential minimum is stable
only if the SM higgs mass satisfies the lower bound constraint \cite{sher}:
\begin{equation}
m_H > 132+2.2(m_t-170)-4.5(\frac{\alpha_s - 0.117}{0.007}),    \label{eq:sher}
\end{equation}
valid for a top mass in the range 160 to 190 GeV,
and
\begin{equation}
m_H > 75+1.64(m_t-140)-3.0(\frac{\alpha_s - 0.117}{0.007}),   \label{eq:sher2}
\end{equation}
valid for a top mass in the range 130 to 150 GeV;
for a top mass between 150 and 160 GeV, approximately 2 GeV must be added to
the bound in Eq.~(\ref{eq:sher2}).  In these equations,
mass units are in GeV, and $\alpha_s$ is the strong coupling constant
at the scale of the $Z$ mass.  These equations are the results of RGE--improved
two--loop calculations, and include radiative corrections to the higgs and
top masses.  They are reliable, and accurate to 1 GeV in the top mass, and
2 GeV in the higgs mass \cite{sher}.

If the universe is allowed to reside in an unstable minimum,
then a similar, but slightly weaker (by $\lsim 5$ GeV for
heavy $m_t$ \cite{sher}) bound results.
The unstable vacuum bound is only slightly weaker because
the instability must be slight to preclude the possibility that
early universe thermal fluctuations pushed the universe into the
wrong but stable vacuum \cite{arnold}.

It has been known for some time \cite{reviews}
that the SM lower bound
rises rapidly as the value of the top mass increases through $M_Z$; below $M_Z$
the bound is of order of the Linde--Weinberg value, $\sim 7$ GeV \cite{LW}.
So what is new here
is the inference from the large reported value for $m_t$ that the SM higgs
lower mass bound dramatically exceeds 100 GeV!
Adding the statistical and systematic errors of the CDF top mass measurement
in quadrature gives a top mass with a single estimated error of
$m_t=174\pm16$ GeV.
The D0 collaboration has used its
nonobservation of top candidates to report a
95\% confidence level lower bound on the top mass of 131 GeV \cite{D0}.
The D0 lower bound is predicated on the presumed dominance of the decay mode
$t\rightarrow b+W$.  The dominance of this mode is supported by the
event signatures in the CDF data.  We will assume the validity of the D0
lower mass bound
\footnote
{A top mass limit independent of the top decay modes is provided
by an analysis of the W boson width: $m_t > 62$ GeV at 95\% confidence
 \cite{Wwidth}.}
{}.
Thus, the D0 lower bound, and the CDF mass value including $1\sigma$
allowances are, respectively, 131, 158, 174, and 190 GeV.
Inputing these top mass values into Eqs.~(\ref{eq:sher}) and~(\ref{eq:sher2})
with $\alpha_s = 0.117$ then yields SM higgs mass lower bounds of
60, 106, 140, and 176 GeV, respectively
\footnote
{We learn here why the LEP experiments have established
the \underline{non--existence}
of the SM higgs particle below a mass value of 64 GeV \cite{LEP}:
when fed into the vacuum stability argument,
the heaviness of the top mass \underline{requires} a low energy
SM higgs desert!}
{}.

This lower limit on the SM higgs from the vacuum stability argument
is a significant
phenomenological constraint, and it rises linearly with
$m_t$, for $m_t \stackrel{>}{\sim} 100$ GeV.
On the other hand, the upper limit on the lightest MSSM higgs rises
quadratically with $m_t$, also for $m_t \stackrel{>}{\sim} 100$ GeV.
Thus, for very heavy $m_t$, the two bounds will inevitably overlap.
Also, for relatively light $m_t$ the bounds may overlap; e.g.\ we
have just seen that the SM lower bound is 60 GeV for $m_t = 131$ GeV,
whereas for large or small $\tan\beta$ the MSSM upper bound is at least
the Z mass.  However, for $m_t$ heavy, but not too heavy, there may
be no overlap.  If so, the first measurement of the lightest higgs mass
will serve to exclude either the SM higgs sector, or the MSSM higgs sector!
In what follows, we show that in fact for $m_t$ around the value reported
by the CDF collaboration, there is a gap between the SM higgs mass
lower bound and the MSSM upper bound.

The vacuum stability bound on the SM higgs
mass is sensitive to the value of $\alpha_s$.
We have taken $\alpha_s=0.117$ (the central value in the work of \cite{sher})
to produce the bounds displayed in Fig. 1.
The reported LEP central value from event shape analyses
is $\alpha_s(M_Z)=0.124\pm0.005$ \cite{bethke}.
Other LEP analyses, and deep inelastic leptoproduction data extrapolated to
the $M_Z$ scale give lower values, resulting in a world average
of $\alpha_s(M_Z)=0.120\pm0.006\pm0.002$ \cite{who}.
If we use  the generous
value $\alpha_s=0.129$, the lower bound on the SM higgs mass decreases
by about 8 GeV for $m_t>160$ GeV. This will decrease the gap between
the SM higgs lower bound and the MSSM higgs upper bound.
However, a decrease of even this magnitude in the SM lower bound
is compensated by the decrease in the MSSM upper bound due to
two-loop contributions not included in our calculations, but discussed
in section 3.

Since vacuum stability of the SM first breaks down for scalar field
fluctuations on the order of $10^6 - 10^{10}$ GeV \cite{sher},
an implicit assumption in this
SM bound is no new physics below $10^{10}$ GeV.
In particular, the stability bound, calculated with perturbation theory,
is not valid if there is a non--perturbatively large value for
the higgs self--coupling $\lambda$
below $\sim 10^{10}$ GeV.
However, if there is a non--perturbatively large value for $\lambda$
below $10^{10}$ GeV, then there will be a Landau pole near or below
$10^{10}$ GeV, which in turn implies a
{\it triviality \underline{lower} bound} on the SM higgs mass of
about 210 GeV.
Before we show how this lower SM bound comes about,
let us state the immediate consequence:
{\it assuming no new fields with mass scales below $10^{10}$ GeV,
either the perturbative stability bound is valid for the SM higgs,
or the non--perturbative triviality lower bound is valid.}
The stability bound is the less restrictive, and we assume it in the
subsequent sections of this paper.

We conclude this section by outlining
the origin of this SM triviality \underline{lower} bound,
$m_h \gsim 210$ GeV.
Various perturbative \cite{triv} and non-perturbative \cite{lattice,wilsonrge}
studies have shown that
a nontrivial (meaning non--vanishing low energy self--coupling)
scalar model can only consistently be
defined as a cut--off theory (i.\ e.\ an effective low energy theory that
ignores new physics at and above the cut--off scale).
An analysis of  the one-loop
renormalization group equation for the Higgs coupling already
reveals some of the consequences of the non-asymptotically free
running of the scalar coupling and its trivial (meaning zero, or nearly so)
infra-red fixed point:
\begin{equation}
\mu \frac{d \lambda}{dt} = \frac{3}{2 \pi^2} \lambda^2
\label{rge}
\end{equation}
The solution to this equation can be written in the form
\begin{equation}
\lambda(\mu) =  \frac{\lambda(\mu_0)}
{1 -\frac{3}{2 \pi^2} \ln ( \frac{\mu}{\mu_0}) \lambda(\mu_o)}
\label{rgesoln}
\end{equation}
It is clear that the coupling constant grows with increasing
energy scale.
Extending the solution beyond its perturbative range of validity,
a pole, called a ``Landau pole'',
manifests itself when the denominator is equal to zero.
The energy scale of the pole is therefore
\begin{equation}
\Lambda = \mu_0 \exp^{\frac{2 \pi^2}{3} \frac{1}{\lambda(\mu_0)}}
\label{pole}
\end{equation}
The occurrence of a Landau pole is usually interpreted as the onset
of non--perturbative physics, or other new physics.
A more rigorous treatment would replace Eq. (\ref{rge})
with two RG equations coupling the running of
$\lambda$ and the running of the top quark yukawa coupling.
However, it is known that inclusion of the top quark terms
only slightly alters the solution $\lambda(\mu)$ and the position of the
Landau pole \cite{lindner}.

A given value of the Higgs mass completely specifies the
solution to the renormalization group equation.
In particular, at $\mu_0= m_h$, the value of the higgs mass
\beq
m_h^2 = 2 v^2 \lambda(m_h)
\label{bcond}
\eeq
as determined by the curvature of the effective potential at its minimum,
fixes the boundary condition in Eq.(\ref{rgesoln}) for $\lambda (\mu)$.
Setting the arbitrary scale $\mu_0$ to $m_h$ in Eq.(\ref{pole}),
and using Eq.(\ref{bcond}) to eliminate $\lambda(m_h)$,
one gets the one--to--one relation between the position $\Lambda$
of the Landau pole and the higgs mass $m_h$:
\begin{equation}
m_h^2 \ln \left( \frac{\Lambda^2}{m_h^2} \right) = \frac{8}{3}
	\pi^2 v^2
\label{oneone}
\end{equation}

If the position of the Landau pole is known, then the higgs mass is determined
implicitly by Eq. (\ref{oneone}).
If it is only known that the Landau
pole is {\it above} a certain scale, say ${\bar \Lambda}$,
then since the higgs mass falls with increasing $\Lambda$,
the higgs mass is only known to be {\it below} the value $m_h({\bar \Lambda})$;
this is the ``triviality upper bound''.
On the other hand, if it is only known that the Landau
pole is {\it below} a certain scale ${\bar \Lambda}$,
then since the higgs mass rises with decreasing $\Lambda$,
the higgs mass is only known to be {\it above} the value $m_h({\bar \Lambda})$.
Thus, the assumption that the self--coupling becomes
non-perturbative below a specific energy scale yields
a minimum value for the Higgs mass, the ``triviality lower bound''.
The inverse of the assumption of no new physics below $\sim 10^{10}$ GeV
that underlies the vacuum stability perturbative lower bound
on the higgs mass therefore implies a non--perturbative \underline{lower}
bound.
{}From Eq.(\ref{oneone}) we find that
$m_h(\Lambda=10^{10} \,{\rm GeV}) = 210$ GeV.
This qualitative discussion based on the perturbative renormalization group
equation is corroborated by several non-perturbative studies,
using the lattice \cite{lattice} or Wilson renormalization flows
 \cite{wilsonrge}, and remains valid even if yukawa and gauge couplings are
included.

\section{The lightest higgs in the MSSM}

The spectrum of the higgs sector in the MSSM contains two CP--even
neutral higgses, $h$ and $H$, with $m_h<m_H$ by convention,
one CP--odd neutral higgs
$A$ and a pair of charged higgs $H^{\pm}$. A common
convenience is to parameterize the higgs sector by the mass of the
CP--odd higgs $m_A$ and the vev ratio
$\tan\beta\equiv v_T/v_B$.
These two parameters completely
specify the masses of the higgs particles at tree level
\begin{eqnarray}
m_{H,h}^2= & \half(m_A^2+m_Z^2)\pm\half\sqrt{(m_A^2-m_Z^2)^2c_{2\beta}^2
+(m_A^2+m_Z^2)^2s_{2\beta}^2} \nonumber \\
m_{H^{\pm}}^2= & m_A^2+m_W^2
\label{eq:treeHiggs}
\end{eqnarray}
implying for example that $m_{H^{\pm}}>m_W$, that the upper bound
on the lightest higgs mass is given by
\beq
m_h\leq |\cos(2\beta)| \, M_Z,
\label{eq:tree}
\eeq
that the lightest higgs mass vanishes at tree level if $\tan\beta=1$,
and that the masses $m_H, m_A$, and $m_{H^{\pm}}$
all increase together as any one of them is increased.
However, radiative
corrections strongly modify the tree level predictions in the neutral
 \cite{mt4,neutral,DreesNojiri,diazhaberii} and
charged \cite{ChargedH,DreesNojiri,diazhaberi}
higgs sectors.  Some consequences are that the
charged higgs can be lighter than the $W$ gauge boson \cite{diazhaberi},
that the $\tan\beta=1$ scenario, in which $m_h=0$ at tree level,
is viable due to the possibility of a large radiatively
generated mass \cite{diazhaberii},
and that the upper bound on the
lightest higgs mass is increased by terms proportional to
$m_t^4 \ln(m_{\tilde{t}}/m_t)$,
as advertised in the introduction
\footnote{
Note that in the susy limit, $m_t=m_{\tilde{t}}$ and the fermion and boson
loop contributions cancel each other.  However, in the real world of broken
susy, $m_t\neq m_{\tilde{t}}$, and
the cancellation is incomplete.
The top quark gets its mass from its
yukawa coupling to the electroweak vev, whereas the scalar top mass arises from
three sources, from D--terms, from the top yukawa coupling, but mainly from
the insertion into the model of dimensionful soft susy--breaking
parameters.  The interplay of these diverse masses
leads to the dramatic correction.
Note that the correction grows logarithmically
as $m_{\tilde{t}}$ gets heavy, rather than decoupling!
For heavy $m_{\tilde{t}}$ the large logarithms can be summed to all orders
in perturbation theory using renormalization group techniques.
Interestingly, the effect is to \underline{lower}
the MSSM upper bound \cite{haberralf}.
}
\cite{mt4}.

An important mechanism for the production of the neutral higgses
in $e^+e^-$ colliders is the brehmsstrahlung of a higgs by a $Z$ gauge boson.
Relative to the coupling of
the SM higgs to two $Z$ bosons, the $ZZH$ coupling is
$\cos(\beta-\alpha)$ and the $ZZh$ coupling is $\sin(\beta-\alpha)$,
where $\alpha$ is the mixing angle in the CP-even neutral higgs mass matrix.
The angle is restricted to $-\frac{\pi}{2}\leq \alpha \leq 0$, and is
given at tree level by
\begin{equation}
\tan2\alpha={{(m_A^2+m_Z^2)}\over{(m_A^2-m_Z^2)}}\tan2\beta.
\label{eq:tantwoalpha}
\end{equation}
{}From Eq.(\ref{tantwoalpha}) it is seen that
the limit $m_A\rightarrow\infty$ is important for three reasons. First,
it requires
$\alpha\rightarrow\beta-\pi/2$, implying that
$\cos(\beta-\alpha)\rightarrow 0$,
\ie, the heavy higgs decouples from the $Z$ gauge boson.
Secondly, it requires
that $\sin(\beta-\alpha)\rightarrow 1$, \ie, the light higgs behaves
like the SM higgs.
And thirdly, $m_A\rightarrow \infty$ is the limit in which
the tree level $m_h$ saturates its maximal value given in Eq. (\ref{eq:tree})
for any value of $\tan\beta$.

We calculate the one-loop corrected lightest MSSM higgs mass,
$m_h$ \cite{marco}.
Included are the full one--loop corrections from the
top/bottom quarks and squarks,
and the leading--log corrections from
the remaining fields (charginos, neutralinos,
gauge bosons, and higgs bosons).
Recently, full one--loop corrections from all particles
 \cite{1loop} have been calculated.
Since the dominant corrections are due to the heavy quarks and squarks,
full one--loop corrections from charginos, neutralinos, gauge and higgs bosons
are well approximated by their leading logarithm terms used here.
Two--loop corrections have recently been calculated also \cite{2loop},
for the limit $\tan\beta \rightarrow \infty$.
Keeping only the leading $m_t$ terms, these corrections have been extrapolated
to all $\tan\beta$.  The graphical result in ref. \cite{2loop}
shows a \underline{lowering} of the MSSM upper bound by several GeV
\footnote
{In ref. \cite{eq} were found small and positive two-loop contributions of the
order $m_t^6$; however, the QCD two-loop contributions found in
ref. \cite{2loop} are
of order $\alpha_s^2m_t^4$, are negative, and dominate the previous ones.
The net effect is to lower the higgs mass bound.
}.
{}From this work \cite {2loop}, we estimate the gap to be wider by several GeV
than the one--loop separation we show in Fig. 1.
This widening further enables a higgs mass measurement to distinguish
the SM and MSSM models.

We choose $m_A$ and all squark mass parameters to be large, approximately
1 TeV
\footnote
{We note that $\lsim 1$ TeV emerges naturally for the heavier superparticle
masses when the MSSM is embedded into a GUT \cite{llabf,kkrw,lnpwz}.
},
in order to find the maximum light higgs mass.
With respect to the squark mixing, we work in two extreme scenarios:\\
(a) no mixing,
\ie, $\mu=A_t=A_b=0$, where $\mu$ is the supersymmetric higgs mass
parameter and $A_i$, $i=t,b$ are the trilinear soft supersymmetry breaking
terms; and\\
(b) maximal mixing with $\mu=A_t=A_b=1$ TeV.\\
The resulting lightest higgs mass as a
function of $\tan\beta$ is shown in Fig. 1 for the four experimentally
motivated
values of the top quark mass discussed earlier.
For the case $\tan\beta\sim 1$, the SM lower bound and the MSSM upper bound
are already non--overlapping at $m_t=131$ GeV\@.
However, for larger $\tan\beta$ values, the overlap persists until
$m_t \stackrel{>}{\sim} 160$ GeV\@.
For the preferred CDF value of $m_t=174$GeV, the gap is present for all
$\tan\beta$, allowing discrimination between the SM and the MSSM based
on the lightest higgs mass alone.  At $m_t=190$ GeV the gap is still widening,
showing no signs of the eventual gap--closure at still higher $m_t$.

Also in Fig. 1 we see that scenario (b) offers a larger
value for the $m_h$ maximum than does scenario (a), except for
the region $\tan\beta\gg 1$. The reason is that among the additional
light higgs mass terms in (b) is a negative term proportional
to $-\mu^4m_b^4/c_{\beta}^4$, which becomes large \cite{haberralf}
when $\tan\beta\gg 1$.
More significant is the fact that the extreme values
in (a) and (b) yield a very similar upper bound in the region of acceptable
$\tan\beta$ values, thereby suggesting insensitivity of the MSSM upper
bound to a considerable range of the squark mixing parameters.

It is known that the branching ratio $B(b\rightarrow s\gamma)$ has
a strong dependence on the susy higgs
parameters \cite{bsfSUSY,diazbsfi}.
However, when all squarks are heavy, as here, the contribution
from the chargino/squark loops to $B(b\rightarrow s\gamma)$
is suppressed. In the case of heavy squarks, the
charged-higgs/top-quark loop may seriously alter the rate, and
strong constraints on the charged higgs minimum
mass result \cite{hewettBBP,diazbsfi}.
This constraint does not affect the present work,
where we take $m_A$ and therefore $m_{H^{\pm}}$ and $m_H$ large
in order to establish the light higgs upper bound:
in the large $m_A$, large squark mass limit,
the ratio $B(b\rightarrow s\gamma)$ approaches the SM value,
consistent with the CLEO bound \cite{expbsf}.

\section{The lightest higgs in non-standard susy models}

The MSSM can be extended in a
straightforward fashion by adding an $SU(2)$ singlet $S$ with vanishing
hypercharge to the theory \cite{Ell}.
As a consequence, the particle spectrum contains an
additional scalar, pseudoscalar, and neutralino.
This extended model, the so--called (M+1)SSM, features four possible
additional terms in the superpotential.
Two of these terms,  $\lambda S H_B \epsilon H_T$ and $\frac{1}{3}\kappa S^3$,
enter into the calculation of the lightest higgs mass; $\epsilon$ is the usual
antisymmetric 2 by 2 matrix.

A tree--level analysis of the eigenvalues
of the scalar mass matrix yields an upper bound on the mass of the
lightest higgs boson:
\begin{equation}
m_h^2 \leq M_Z^2 \left\{ \cos^2 2 \beta +2 \frac{\lambda^2}{g_1^2 + g_2^2}
\sin^2 2 \beta \right\}.
\end{equation}
The first term on the right hand side
is just the MSSM result of Eq. (\ref{eq:tree}).
The second term is positive semidefinite, and so weakens the bound
compared to its counterpart in the MSSM.
Moreover, the parameter $\lambda$ is {\it a priori} free, and so the second
term may \underline{considerably}
weaken the upper bound \cite{Vel,nonVel,Kan}.
However, there are two
cases where the bound will suffer only a minor adjustment.
The first is the large $\tan\beta$ scenario, where $\cos^2 2\beta$ is
necessarily $\gg \sin^2 2\beta$.  The second is
when the theory is embedded into a GUT.  In this case,
the strength of $\lambda$ at the susy--breaking scale, $M_{SUSY}, $ is limited:
even if $\lambda$ assumes a high value at the GUT scale,
the nature of the
renormalization group equations is such that
its evolved value at the susy--breaking scale
is a rather low, pseudo--fixed point.
Under the assumption that all coupling constants remain perturbative
up to the GUT scale, it is therefore possible to calculate a maximum value
for the mass of the lightest higgs boson \cite{Vel,nonVel}.
It turns out that this lightest mass upper bound occurs when $\kappa$ is
close to zero.
The higgs mass upper bound depends on the value
of the top yukawa $g_t$ at the GUT scale through the renormalization group
equations.  Above $M_{SUSY}$ the running of the coupling
constants is described by the (M+1)SSM renormalization group
equations, whereas below this scale the SM renormalization group equations
are valid. At $M_{SUSY}$ the boundary conditions
\begin{eqnarray}
\lambda^{SM} & = & \frac{1}{8} \left( g_1^2 +g_2^2 \right)
\left( \cos^2 2 \beta + 2 \frac{\lambda^2}
{ g_1^2 + g_2^2 } \sin^2 2 \beta \right),
\nonumber \\
g_t^{SM} & = & g_t \sin \beta,
\label{eq:rge}
\end{eqnarray}
incorporate the transition from the (M+1)SSM to the SM. Here
$\lambda^{SM}$ and $g_t^{SM}$ are the standard model higgs self coupling
and top quark yukawa coupling respectively. The value of the
higgs boson mass is determined implicitly by the equation
$2 \lambda^{SM} \left( m_h \right) v_{SM}^2 = m_h^2$.
This RGE procedure of running couplings from $M_{SUSY}$ down
takes into account logarithmic radiative corrections to
the higgs boson mass, in particular those caused by the heavy top quark.

In Fig. 2 we show the maximum value of the
higgs boson mass as a function of $\tan \beta$ for
the chosen values of the top quark mass $m_t$.
We have adopted a susy--breaking
scale of $M_{SUSY}=1\ TeV$; this value is consistent with the notion of
stabilizing  the weak--to--susy GUT hierarchy, and is the value favored by RGE
analyses of the observables $\sin^2\theta_W$ and $m_b/m_{\tau}$.
The bounds in Fig. 2 are quite insensitive to the choice of $M_{SUSY}$,
increasing very slowly
as $M_{SUSY}$ increases \cite{Vel}.
We have assumed that all superpartners and all higgs bosons except
for the lightest one are heavy, \ie $\sim M_{SUSY}$.
In Fig. 2 it is revealed that
for low values of the top quark mass ($\sim M_Z$), the mass upper bound on the
higgs boson in the
(M+1)SSM will be substantially higher than in the MSSM at
$\tan\beta \lsim$ a few. This is because $\lambda(m_h)$ is large for low $m_t$,
and because $\sin^2 2\beta \gsim \cos^2 2\beta$ for $\tan\beta \lsim$ a few.
However, for a larger top quark mass the difference
between the MSSM and (M+1)SSM upper bounds diminishes.
This is because $\lambda(m_h)$ falls with increasing $m_t$, and because
there is an increasing minimum value for
$\sin\beta=g_t^{SM}/g_t$ [from the second of Eqs. (\ref{eq:rge})],
and therefore for $\tan\beta$,
when $m_t\propto g_t^{SM}$ is raised and $g_t$ is held to be perturbatively
small up to the GUT scale.
This increasing minimum value of $\tan\beta$ is evident in the curves of
Fig. 2.
A comparison of Figs. 1 and 2 reveals that the (M+1)SSM and MSSM bounds are
very similar at $\tan\beta\gsim 6$.
For $m_t$ at or above the CDF value, only this  $\tan\beta\gsim 6$
region is viable in the (M+1)SSM model.
Since the (M+1)SSM model was originally constructed to test the robustness of
the MSSM, it is gratifying that the two models show a very similar upper bound.

The results for more complicated extensions of the minimal model tend
to be similar \cite{Kan}. In general, the mass of the lightest higgs boson
at tree level is limited by $M_Z$ times a factor proportional to the
dimensionless coupling constants in the higgs sector.
The requirement
of perturbative unification restricts the  value of these coupling
constants at the electroweak scale, and the maximum value of the lightest higgs
boson mass is therefore never much larger than $M_Z$.

We have seen that the SM, MSSM, and the (M+1)SSM electroweak models can be
disfavored or ruled out by a measurement of $m_h$;
and that a ``forbidden'' mass gap exists for $m_t\gsim 160$ GeV.
We next give an example of
a non-standard susy model that cannot be embedded in a GUT,
and requires a low susy breaking scale:
a gauged, non--linear, supersymmetric sigma model.
The simplest supersymmetric model with a non-linear
representation of the $SU(2) \times U(1)$ symmetry is
obtained by imposing the constraint $H_T \epsilon H_B = \frac{1}{4} v_{SM}^2
\sin^2 2 \beta$ on the action of the MSSM \cite{Fer}.
This constraint is the only one possible in the MSSM higgs sector that
obeys supersymmetry, is invariant under $SU(2)\times U(1)$,
and leaves the vev in a global minimum
\footnote
{This MSSM non--linear sigma model is not the formal heavy higgs limit of
the MSSM.  This is in contrast to the non--linear sigma
models that result from the heavy higgs limit of the SM,
or of the (M+1)SSM.  The difference is that MSSM does
not contain an independent, dimensionless, quartic
coupling constant $\lambda$ in the higgs sector which can be taken to infinity,
whereas the SM and (M+1)SSM do.
}.
As a result of this
constraint one of the scalar higgs bosons, the pseudoscalar, and one of
the neutralinos are eliminated from the particle spectrum. The remaining higgs
boson has a mass
$m_h^2 =  M_Z^2 + (\hat{m}_T^2 + \hat{m}_B^2) \sin^2 2 \beta$, and the charged
higgs bosons have masses
$m_{H^{\pm}}^2 = M_W^2 + (\hat{m}_T^2 + \hat{m}_B^2)$.
Here, $\hat{m}_T^2$ and $\hat{m}_B^2$
are soft, dimensionful, susy--breaking terms; they may be positive or negative.

In order for the
notion of a supersymmetric non-linear model to be relevant, the susy
breaking scale is required to be much smaller than the chiral
symmetry breaking scale $4 \pi v^{SM}$. The natural magnitude
for the parameters
$\hat{m}_B^2$ and $\hat{m}_T^2$
is therefore of the order of $M_Z^2$. Consequently,
both the neutral and the charged higgs bosons have masses of at most a
few multiples of $M_Z$ in the non--linear model.
This formalism of the effective action allows a description of the low energy
physics \underline{independent} of the particular strongly--interacting
underlying theory from which it derives.
Thus we believe that the non--linear MSSM model presented here is probably
representative of a class of underlying strongly--interacting susy models.
The lesson learned then is that measuring a
value for $m_h$ at $\lsim 300$ GeV cannot validate the SM, MSSM, (M+1)SSM,
or any other electroweak model.  However, the premise of this present article
remains valid, that
such a measurement should rule out one or more of these popular models.

\section{Supersymmetric Grand Unified Theories}

Supersymmetric grand--unified theories (susy GUTs)
are the only simple models in which\\
(i) the three low energy gauge
coupling constants are known to merge at the GUT scale;\\
(ii) the correct low energy value for the
weak mixing angle $\sin\theta_W$ is obtained;\\
(iii) hierarchy and parameter--naturalness issues are solved.\\
Thus, it is well motivated to consider the grand unification of the low
energy susy models.
Many susy GUTs reduce at low energies to the MSSM
with additional constraints on the parameters \cite{kkrw}.
The additional constraints must yield an effective low energy theory
that is a special case of the general MSSM we have just considered.
Accordingly, the upper limit
\footnote
{In fact, the additional restrictions may be so constraining as to also
yield a {\it lower} limit on the lightest higgs mass,
in addition to the upper limit.
For example, $m_h> 85$ GeV for $\tan\beta > 5$ and $m_t=170$ GeV is reported in
ref.\cite{kkrw}, and a similar result is given in \cite{lnpwz}.
}
on $m_h$ in such susy GUTs is
in general \underline{more restrictive}
than the bound presented in section 3.
The assumption of gauge coupling constant unification
(with its implied desert between $M_{SUSY}$ and $M_{GUT}$) presents no
significant constraints on the low energy MSSM parameters \cite{kkrw,lang1}.
However, the further assumption that the
top yukawa coupling remains perturbatively small up to $M_{GUT}$
leads to the low energy constraint $0.96 \leq \tan\beta$.  This is because
the RGE evolves a large but perturbative top yukawa coupling
at $M_{GUT}$ down to its well--known infrared pseudo--fixed--point
value at $M_{SUSY}$ and below, resulting in the top mass value
$\sim 200 \sin\beta$ GeV.
If the bottom yukawa is also required to remain perturbatively small
up to $M_{GUT}$, then $\tan\beta \leq 52$ \cite{bbk} emerges as a second
low energy constraint.

The pseudo--fixed--point solution is not a true fixed--point,
but rather is the low energy yukawa value that runs to become a Landau pole
(an extrapolated singularity, presumably tamed by new physics)
near the GUT scale.
The apparent CDF top mass value is within the estimated range of the
pseudo--fixed--point value.  Thus it is attractive to assume
the pseudo--fixed--point solution.
With the additional assumptions
that the electroweak symmetry is radiatively broken \cite{drees}
(for which the magnitude of the top mass is crucial) and
that the low energy MSSM spectrum is defined by a small number of
parameters
at the GUT scale (the susy higgs mass parameter $\mu$ -- which is also
the higgsino mass, and four universal soft susy--breaking mass
parameters: the scalar mass, the bilinear and trilinear masses, and the
gaugino mass),
two compact, disparate ranges for $\tan\beta$ emerge:
$1.0\leq \tan\beta \leq 1.4$ \cite{bbk},
and a large $\tan\beta$ solution $\sim m_t/m_b$.
Reference to our Figs. 1 and 2 shows that the gap between the SM and MSSM
is maximized in the small $\tan\beta$ region and minimized in the large
$\tan\beta$ region, whereas just the opposite is true for the SM and
(M+1)SSM models.  Moreover, the (M+1)SSM model is an inconsistent theory
in the small $\tan\beta$ region if $m_t \gsim$ 160 GeV.

In fact, a highly constrained
low $\tan\beta$ region $\sim 1$ and high $\tan\beta$ region
$\gsim$ 40--70
also emerge when
bottom--$\tau$ yukawa unification at the GUT scale is imposed on
the radiatively broken model \cite{fixpt,barger,cw,lang2,copw,krwk}.
Bottom--$\tau$ yukawa coupling unification is attractive in that it is
natural in susy SU(5), SO(10), and $E_6$,
and explains the low energy relation, $m_b\sim 3 m_{\tau}$.
With bottom-$\tau$ unification,
the low to moderate $\tan\beta$ region requires
the proximity of the top mass to its fixed--point value \cite{bcpw},
while the high $\tan\beta$ region also requires the proximity of the bottom
and $\tau$ yukawas to their fixed--point; the emergence of the two $\tan\beta$
regions results from these two possible ways of assigning fixed--points.

The net effect of the yukawa--unification constraint in susy GUTs is
necessarily to widen the mass gap
between the light higgs MSSM and the heavier higgs SM,
thus strengthening the potential for experiment to distinguish the models.
The large $\tan\beta$ region is
disfavored by proton stability \cite{arnonath}.
Adoption of the favored low to moderate $\tan\beta$ region
leads to a highly
predictive framework for the higgs and susy particle
spectrum \cite{lang2,copw,krwk}.
In particular, the fixed--point relation $\sin\beta \sim m_t/(200 GeV)$
fixes $\tan\beta$ as a function of $m_t$.
For a heavy top mass as reported by CDF, one has $\tan\beta \sim$ (1, 2)
for $m_t =$ (140, 180) GeV.
Since $\tan\beta \sim 1$ is the value for which
the $m_h$ upper bound is minimized (the tree--level contribution to $m_h$
vanishes),
the top yukawa fixed--point models offer a high likelihood for $h^0$
detection at LEP200.
Reduced $m_h$ upper bounds have been reported in \cite{barger,cw,lang2,krwk}.
These bounds are basicaly our bound
in Fig. 1 for $\tan\beta\gsim 1$, where small differences appear when
different methods and approximations are used.
These reduced bounds are due to the
small $\tan\beta$ restriction, an inevitable consequence of assigning the
top mass, but not the bottom mass, to the pseudo fixed--point.

Even more restrictive susy GUTs have been analyzed.
These include the ``no--scale'' or minimal supergravity models \cite{gp},
in which the soft mass parameters $m_0$ (universal scalar mass)
and $A$ are zero at the GUT
scale; and its near relative, the superstring GUT,
in which the dilaton vev provides the dominant source of susy
breaking and so $m_0$, $A$, and the gaugino mass parameter all scale
together at the GUT scale \cite{bgkp}.
Each additional constraint serves to further widen the SM/MSSM higgs
mass gap.

In radiatively broken susy GUTs with universal
soft parameters, the superparticle spectrum emerges at $\lsim$ 1 TeV.
If the spectrum in fact saturates the 1 TeV value, then as we have seen
the Feynman rules connecting $h^0$ to SM particles
are indistinguishable from the Feynman rules of the SM higgs.
Thus, it appears that
if a susy GUT is the choice of Nature, then the mass of the lightest higgs,
but not the higgs production or dominant decay modes,
may provide our first hint of grand unification.

\section{Discovery potential for the higgs boson}

The higgs discovery potential of LEPII \cite{gunion,djouadi}
depends on the energy at
which the machine is run.  A higgs mass
up to 105 GeV is detectable at LEPII with the $\sqrt{s}=200$ GeV option
(LEP200), while a higgs mass only up to 80 GeV is detectable with LEP178.
As we have shown,
the large value of $m_t$ reported by CDF raises the upper limit
on the MSSM $h^0$ mass to $\sim 130$ GeV.
Near this limit the MSSM higgs has the production and decay properties of
the SM higgs.  Discovery of this
lightest MSSM higgs then argues strongly for the LEP200 option over LEP178.
Furthermore, for any choices of the MSSM parameters,
associated production of either $h^0 Z$ or $h^0 A$ is guaranteed
at LEP200 as long as $m_{\tilde{t}}\lsim$ 300 GeV \cite{gunion}.
Even better would be LEP230, where detection of $Z h^0$ is guaranteed
as long as $m_{\tilde{t}}\lsim$ 1 TeV \cite{gunion}.
At an NLC300 (the Next Linear Collider), detection of $Z h^0$ is guaranteed
for MSSM or for (M+1)SSM \cite{gunion}.
Turning to hadron colliders \cite{zerwas,mrenna}, it is
now believed that while the higgs cannot be discovered
at Fermilab's Tevatron with its present energy and luminosity,
the mass range 80 GeV to 130 GeV is detectable at any hadron collider
with $\sqrt{s} \gsim 2$ TeV
and an integrated luminosity ${\cal L} \gsim 10 {\rm fb}^{-1}$ \cite{mrenna};
the observable mass window widens significantly with increasing luminosity,
but very little with increasing energy.
For brevity, we will refer to this High Luminosity DiTevatron hadron machine
as the ``HLDT''.
If the SM desert ends not too far above the electroweak scale, then the
SM higgs may be as heavy as $\sim$ 600--800 GeV
\footnote
{Theorists would prefer an even lower value of $\lsim 400$ GeV,
so that perturbative calculations in the SM converge \cite{durand}.
}
(but not heavier, according to the triviality argument),
in which case only the LHC (and not even NLC500) guarantees detection.

We present our conclusions on detectability for each of the four $m_t$ values
that we have considered
\footnote
{Recall that the SM vacuum stability bound assumes
a dessert up to at least $10^{10}$ GeV}
:\\
(i) if $m_t \sim 131$ GeV, then
the SM higgs mass lower bound from vacuum stability
is 60 GeV; a SM mass up to (80,105,130) GeV is
detectable at (LEP178,LEP200,HLDT);
and the MSSM $h^0$ is
certainly detectable at LEP178 for $\tan\beta \sim$ 1--2, and
certainly detectable at LEP200 for all $\tan\beta$.\\
(ii) if $m_t \sim 158$ GeV, then
the SM lower bound rises above 100 GeV,
so the SM higgs cannot be detected at LEP178 or LEP200,
but is still detectable at the HLDT if its mass is below 130 GeV;
the lightest MSSM higgs is certainly detectable at LEP178 if $\tan\beta$ is
very close to 1, and is certainly detectable at LEP200 if $\tan\beta$
is $\lsim 3$.\\
(iii) if $m_t \sim 174$ GeV, then
the SM higgs is above 140 GeV, out of reach for LEPII and the HLDT;
the MSSM higgs is certainly detectable at LEP200 if $\tan\beta \sim$ 1--2.\\
(iv) if $m_t \sim 190$ GeV, then
the SM higgs is above 176 GeV in mass;
at any $\tan\beta$ value, the MSSM higgs is not guaranteed to be
detectable at LEP200,
but is certainly detectable at the HLDT if $\tan\beta \sim$ 1--3.\\
It is interesting that the $h^0$ mass range is most accessible to experiment
if $\tan\beta \sim$ 1--3, just the parameter range favored by susy GUTs.

\section{Discussion and conclusions}

We repeat that the lightest MSSM higgs is guaranteed detectable at LEP230;
and that the lightest (M+1)SSM higgs and MSSM higgs
are guaranteed detectable at a NLC300 and at the LHC.
Since there is no lower bound on the lightest MSSM higgs mass other than the
experimental bound,
the MSSM $h^0$ is possibly detectable even at LEP178 for all $\tan\beta$,
but there is no guarantee.
The SM higgs is guaranteed detectable only at the LHC;
if $m_t \sim 174$ GeV, then the SM higgs will not be produced until the LHC
or NLC is available.
Thus, one simple conclusion is that LEPII has a tremendous potential to
distinguish MSSM and (M+1)SSM symmetry
breaking from SM symmetry breaking.

It is worth noting that with enough higgs events, measurement of
certain rare decay modes is very sensitive to non--SM higgs physics.
For example, modes forbidden at tree--level but induced at one loop,
such as $h\rightarrow \gamma \gamma$, $h\rightarrow \gamma Z$,
and $Z \rightarrow \gamma h$,
receive comparable contributions from standard and superpartner
particles in the loop.  The branching fractions for these modes
may vary by an order of magnitude or more from the SM values \cite{yw}.
However, measurements
of these rare modes will require the LHC or the photon--photon
collider option of the NLC500 \cite{mrenna}.

Thus, either the direct detection of the lightest higgs particle as discussed
herein or measurements of rare higgs decays have the potential to
distinguish the SM and MSSM symmetry breaking sectors.
The mass measurement will come first.
We have shown that for a top quark mass $\sim 174$ GeV, as reported by CDF,
a gap exists between the SM higgs mass ($\gsim 140$ GeV) and the lightest
MSSM higgs mass ($\lsim 130$ GeV).  Thus, the first higgs mass measurement
will eliminate one of these popular models.
Most of the MSSM mass range is accessible to LEPII.
If a higgs is discovered
at LEPII, the SM higgs sector is ruled out.
For the (M+1)SSM with the assumption of perturbative unification,
conclusions remain the same as for the MSSM.
\vspace{1.0cm}
\\
{\bf Acknowledgements:}\\
This work was supported in part by the U.S. Department of Energy grant
no.\ DE-FG05-85ER40226, and the Texas National Research Laboratory Commission
grant no.\ RGFY93--303.
\vfill\eject

\vfill\eject
%
%
%
{\bf Figure Captions:}\\
\noindent
{\bf Fig. 1} The solid curves reveal the
upper bound on the lightest MSSM higgs particle
vs.\ $\tan\beta$, for top mass values of (a) 131 GeV,
(b) 158 GeV, (c) 174 GeV, and (d) 190 GeV.
Two extreme choices of susy parameters are invoked:
the higher curve is for $\mu=A_t=A_b=1$ TeV, and the lower curve is for
$\mu=A_t=A_b=0$;
in both cases, $m_A=m_{\tilde{q}}=1$ TeV and $m_b(M_Z)=4$ GeV are assumed.
The dashed curve is the ($\tan\beta$ independent) lower bound on the SM
higgs mass such that the universe sits in the SM vacuum
since the time of the electroweak phase transition.
\\
\noindent
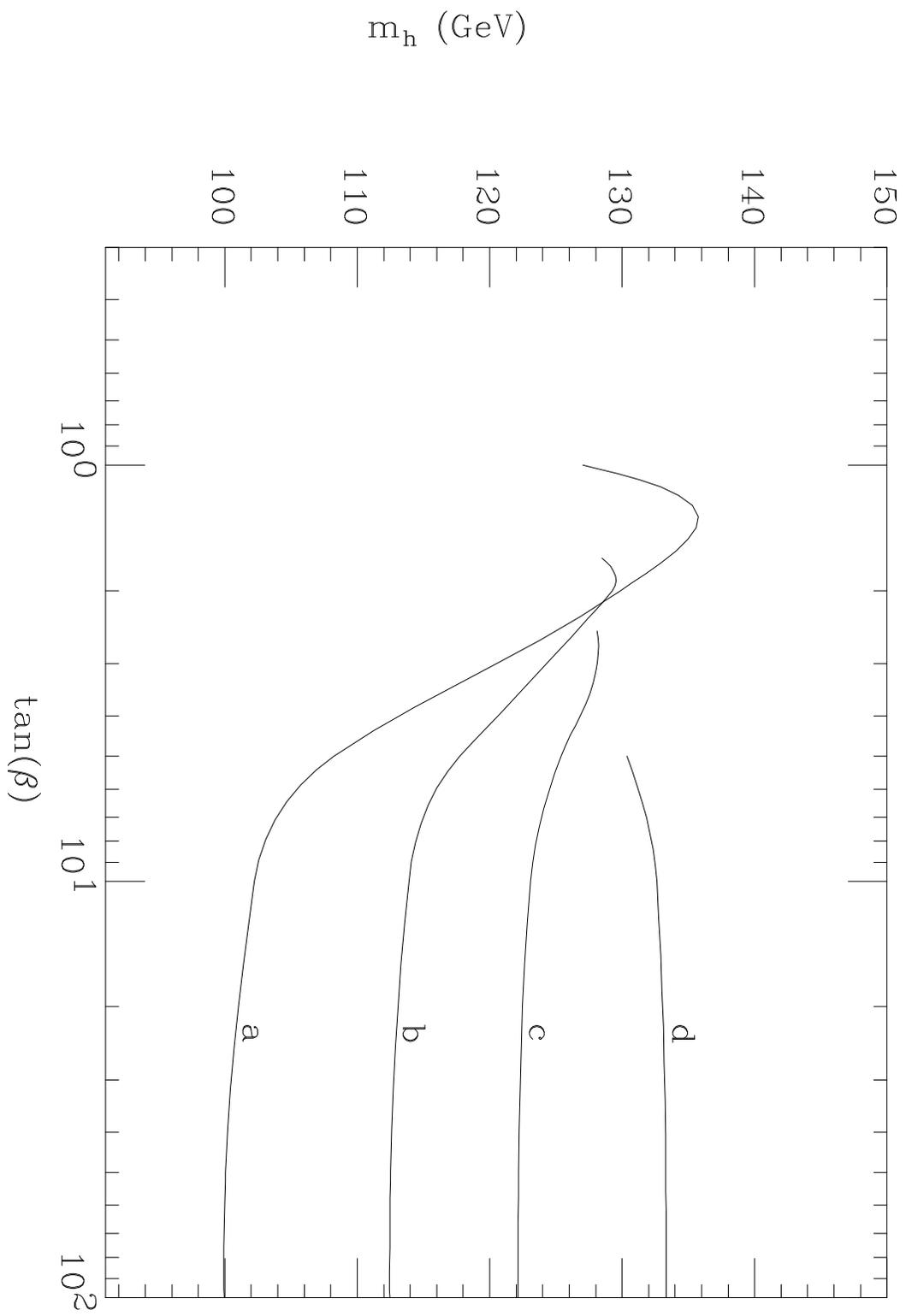
{\bf Fig. 2} Upper bound on the lightest (M+1)SSM higgs vs.\ $\tan\beta$,
for the top mass values (a) 131 GeV,
(b) 158 GeV, (c) 174 GeV, and (d) 190 GeV.
All superparticles and higgses beyond the lightest are assumed to be heavy,
at the chosen susy--breaking scale of 1 TeV. The GUT scale is taken as
$10^{16}$ GeV.
\\
\end{document}